\documentclass[10pt, conference, compsocconf]{IEEEtran}
\pdfoutput=1

\usepackage{url}
\usepackage{cite}
\usepackage{amsmath,amssymb,amsfonts}
\usepackage{algorithmic}
\usepackage{graphicx}
\usepackage{textcomp}
\usepackage{paralist}
\usepackage{listings}
\usepackage{booktabs}
\usepackage{hhline}
\usepackage{framed,multirow}
\usepackage{soul}
\usepackage{cancel}
\usepackage{booktabs}
\usepackage{latexsym}
\usepackage{array}
\usepackage{stackengine}
\usepackage{relsize}
\usepackage[inline]{enumitem}
\usepackage[skip=2pt]{subcaption}
\usepackage[ruled,vlined]{algorithm2e}

\SetCommentSty{mycommfont}
\usepackage[dvipsnames]{xcolor}

\usepackage{url}
\usepackage{hyperref}
\usepackage{glossaries}

\hypersetup{
    colorlinks=true,
    linkcolor=blue,
    filecolor=magenta,      
    urlcolor=cyan,
    pdftitle=WholeSIGHT_PCa_Segmentation_Classification,
    pdfinfo={
        Title={Weakly supervised joint whole-slide segmentation and classification in prostate cancer},
        Author={Pushpak Pati, Guillaume Jaume, Zeineb Ayadi, Kevin Thandiackal, Behzad Bozorgtabar, Maria Gabrani, Orcun Goksel},
      }
}
\definecolor{newcolor}{rgb}{.8,.349,.1}

\newcommand{\ie}{\textit{i}.\textit{e}., }
\newcommand{\eg}{\textit{e}.\textit{g}., }

\newcommand{\gwsss}{\textsc{WholeSIGHT}}
\newacronym{wsss}{WSSS}{weakly-supervised semantic segmentation}
\newacronym{tg}{TG}{tissue-graph}

\newcommand{\mlp}{\mathrm{MLP}}

\newcommand{\concat}{\mathrm{CONCAT}}
\newcommand{\gradcam}{\textsc{Grad-CAM}}
\newcommand{\graphgradcam}{\textsc{GraphGrad-CAM}}

\newcommand{\aggregate}{\textsc{Aggregate}}
\newcommand{\update}{\textsc{Update}}
\newcommand{\readout}{\textsc{Readout}}

\newcommand{\real}{\mathbb{R}}
\newcommand{\std}[1]{{\scriptsize$\pm$#1}}

\newacronym[plural=WSIs,firstplural=Whole-Slide Images (WSIs)]{wsi}{WSI}{whole-slide image}
\newacronym[plural=TMAs,firstplural=Tissue-Micro Array (TMAs)]{tma}{TMA}{Tissue-Micro Array}
\newacronym{he}{H\&E}{Hematoxylin \& Eosin}
\newacronym{cp}{CompPath}{Computational Pathology}
\newacronym{dp}{DigPath}{Digital Pathology}

\newacronym{ml}{ML}{Machine Learning}
\newacronym{nlp}{NLP}{Natural Language Processing}
\newacronym{cv}{CV}{Computer Vision}
\newacronym{xai}{XAI}{Explainable AI}
\newacronym{dl}{DL}{Deep Learning}
\newacronym{ai}{AI}{Artificial Intelligence}
\newacronym{cad}{CAD}{Computer-Aided Diagnosis}
\newacronym[plural=CNNs,firstplural=Convolutional Neural Networks (CNNs)]{cnn}{CNN}{Convolutional Neural Network}
\newacronym[plural=RNNs,firstplural=Recurrent Neural Networks (RNNs)]{rnn}{RNN}{Recurrent Neural Network}
\newacronym[plural=MLPs,firstplural=Multi-Layer Perceptrons (MLPs)]{mlp}{MLP}{Multi-layer Perceptron}
\newacronym[plural=GNNs,firstplural=Graph Neural Networks (GNNs)]{gnn}{GNN}{Graph Neural Network}
\newacronym[plural=GCNs,firstplural=Graph Convolutional Networks (GCNs)]{gcn}{GCN}{Graph Convolutional Network}
\newacronym[plural=MPNNs,firstplural=Message Passing Neural Networks (MPNNs)]{mpnn}{MPNN}{Message Passing Neural Network}
\newacronym{mil}{MIL}{Multiple Instance Learning}
\newacronym{gin}{GIN}{Graph Isomorphism Network} 
\newacronym{pna}{PNA}{Principal Neighborhood Aggregation}
\newacronym{cam}{CAM}{Class Activation Map}
\newacronym{gap}{GAP}{Global Average Pooling}
\newacronym{wsight}{WholeSIGHT}{Whole-slide SegmentatIon using Graphs for HisTology}
\newacronym{de}{DE}{Deep Ensemble}
\newacronym{vae}{VAE}{Variational Auto-Encoder}
\newacronym{em}{EM}{Expectation Maximization}
\newacronym{mle}{MLE}{Maximum Likelihood Estimation}
\newacronym{nll}{NLL}{Negative Log-Likelihood}
\newacronym{sgd}{SGD}{Stochastic Gradient Descent}
\newacronym{relu}{ReLU}{Rectified Linear Unit}
\newacronym{ece}{ECE}{Expected Calibration Error}
\newacronym{mmce}{MMCE}{ Maximum Mean Calibration Error}
\newacronym{tsne}{t-SNE}{t-distributed stochastic neighbor}

\newacronym{knn}{k-NN}{k-Nearest Neighbors}
\newacronym[plural=RAGs,firstplural=Region Adjacency Graphs (RAGs)]{rag}{RAG}{Region Adjacency Graph}

\newacronym[plural=RoIs,firstplural=Regions-of-Interest (ROIs)]{roi}{RoI}{Region-of-Interest}
\newacronym[plural=TRoIs,firstplural=Tumor RoIs (TRoIs)]{troi}{TRoI}{Tumor RoI}
\newacronym{slic}{SLIC}{Simple Linear Iterative Clustering}

\newacronym[plural=CPUs,firstplural=Central Processing Unit (CPUs)]{cpu}{CPU}{Central Processing Unit}
\newacronym[plural=GPUs,firstplural=Graphics Processing Unit (GPUs)]{gpu}{GPU}{Graphics Processing Unit}
\newacronym{ui}{UI}{User Interface}
\newacronym{api}{API}{Application Programming Interface}
\newacronym{dgl}{DGL}{Deep Graph Library}

\begin{document}
\title{Weakly Supervised Joint Whole-Slide Segmentation and Classification \\ in Prostate Cancer}

\author{
\IEEEauthorblockN{
Pushpak Pati$^{\dagger 1 *}$,
Guillaume Jaume$^{\dagger 2,3,4,5}$, 
Zeineb Ayadi$^8$,
Kevin Thandiackal$^{1,6}$, \\
Behzad Bozorgtabar$^7$,
Maria Gabrani$^1$, 
Orcun Goksel$^{6,9}$
}
\IEEEauthorblockA{
\\
$^1$IBM Research Europe, Switzerland \\
$^2$Department of Pathology, Brigham and Women's Hospital, Harvard Medical School, USA \\
$^3$Department of Pathology, Massachusetts General Hospital, Harvard Medical School, USA \\
$^4$Cancer Program, Broad Institute of Harvard and MIT, USA \\
$^5$Data Science Program, Dana-Farber/Harvard Cancer Center, USA \\
$^6$Computer-Assisted Applications in Medicine, ETH Zurich, Switzerland \\
$^7$Signal Processing Laboratory 5, EPFL, Switzerland \\
$^8$EPFL, Switzerland \\
$^9$Department of Information Technology, Uppsala University, Sweden}
}

\maketitle
\thispagestyle{plain}
\pagestyle{plain}

\begingroup\renewcommand\thefootnote{*}
\footnotetext{Corresponding author: Pushpak Pati. Email: \hyperlink{pus@zurich.ibm.com}{pus@zurich.ibm.com}}

\begin{abstract}
The segmentation and automatic identification of histological regions of diagnostic interest offer a valuable aid to pathologists. However, segmentation methods are hampered by the difficulty of obtaining pixel-level annotations, which are tedious and expensive to obtain for Whole-Slide images (WSI). To remedy this, weakly supervised methods have been developed to exploit the annotations directly available at the image level. However, to our knowledge, none of these techniques is adapted to deal with WSIs. In this paper, we propose $\gwsss$, a weakly-supervised method, to simultaneously segment and classify WSIs of arbitrary shapes and sizes. Formally, $\gwsss$ first constructs a tissue-graph representation of the WSI, where the nodes and edges depict tissue regions and their interactions, respectively. During training, a graph classification head classifies the WSI and produces node-level pseudo labels via  post-hoc feature attribution. These pseudo labels are then used to train a node classification head for WSI segmentation. During testing, both heads simultaneously render class prediction and segmentation for an input WSI. We evaluated $\gwsss$ on three public prostate cancer WSI datasets. Our method achieved state-of-the-art weakly-supervised segmentation performance on all datasets while resulting in better or comparable classification with respect to state-of-the-art weakly-supervised WSI classification methods.
Additionally, we quantify the generalization capability of our method in terms of segmentation and classification performance, uncertainty estimation, and model calibration. 
\end{abstract}


\IEEEpeerreviewmaketitle

\section{Introduction}
\label{sec:introduction}
Prostate cancer is the second most frequently diagnosed cancer in men in the United States, with 250,000 new registered cases resulting in 35,000 deaths in 2022~\cite{siegel22}. Yet the number of pathologists, whose role is critical in the diagnosis and management of cancer patients, is gradually declining. In the United States, an 18\% decrease was recorded between 2007 and 2017, resulting in a 42\% increase in the average workload~\cite{wilson18}. In addition, the practice of uro-pathology also has its share of challenges~\cite{amin2015update}. Although the diagnostic criteria for grading prostate cancer are established~\cite{tan19}, the continuum of phenotypic features across the diagnostic spectrum leaves room for disparities, with significant intra- and interobserver variability~\cite{gomes14, elmore15}. The manual inspection of slides is also tedious and time-consuming, and would benefit from automation and standardization. These elements justify the development of \gls{cad} tools to automate the diagnostic workflow.

To this end, several \gls{ai} based \gls{cad} tools are proposed, including  nucleus segmentation and classification~\cite{graham19,pati20-media,graham2023one}, gland segmentation~\cite{sirinukunwattana17,binder19,graham2023one}, and tumor detection~\cite{bejnordi17-jama,aresta19}.
Albeit the remarkable performance, these tools often demand task- and tissue-specific annotations on large datasets, which are tedious, time-consuming and often infeasible to acquire. 
For reducing annotation requirements, different approaches are proposed, in particular, weakly-supervised methods based on \gls{mil} framework for the automatic \emph{classification} of \glspl{wsi}~\cite{lu20b,campanella19}. 

Although classification is useful, it remains limited in its role of supporting the pathologist's attention during diagnosis. In this context, semantic segmentation methods are preferable as they enable the generation of pixel-level delineation of the tissue constituents that can highlight diagnostically relevant regions.
Such visualization allows for strengthening trust between pathologists and \gls{cad} tools. Additionally, the identified regions can be leveraged by a classifier to improve patient diagnosis.
However, semantic segmentation generally requires pixel-level labels, which makes it more demanding in terms of annotations than classification tasks. For this reason, the development of \gls{wsss} methods appears as the most adequate response.

While \gls{wsss} has been successful on natural images, it encounters various challenges when applied to histopathology images~\cite{chan2021comprehensive}, as they, (1) contain fine-grained objects with large intra-class variations~\cite{xie19}; (2) often include ambiguous boundaries among histology components~\cite{xu17}; (3) can be several giga-pixels with arbitrary tissue sizes. Nevertheless, some \gls{wsss} methods are proposed for various histology tasks. The methods by~\cite{xu14,hou16,jia17,xucamel19,ho21,zhang2021joint,han2022multi} performing \gls{wsss} at patch-level are limited by the need for patch-level annotations, and inability to perform global contextualized \gls{wsi} segmentation. While~\cite{chan19,silva2021weglenet} scale to larger tiles, they pose high computational complexity and memory requirements for operating on \glspl{wsi}. The methods by \cite{chan19,han2022multi} require \textit{exact} tile annotations for model training, \ie a precise denomination of each lesion type in a tile, which requires pathologists to annotate images beyond standard clinical needs. On a different note, recent \gls{wsi} classification methods use attention mechanisms or feature attributions to highlight salient regions~\cite{lu20b}. Though these regions are informative for visual assessment, they are insufficient, incomplete, and blurry for accurately delineating relevant regions. Additionally, producing granular saliency requires densely overlapping patch predictions, which is computationally expensive while working with \glspl{wsi}.

In view of the aforementioned limitations of the \gls{wsss} methods, we propose $\gwsss$, ``Whole-slide SegmentatIon using Graphs for HisTopathology'', that can simultaneously segment and classify arbitrarily large histopathology images by using \gls{wsi}-level labels, and without any task-specific assumptions or post-processing. Formally, $\gwsss$ transforms an image into a superpixel-based \gls{tg}, and considers the segmentation problem as a \emph{node-classification} task. $\gwsss$ incorporates both local and global tissue microenvironment to perform contextualized segmentation, principally in agreement with inter-pixel relation-based \gls{wsss}~\cite{ahn19}. 
To summarize, our contributions are:
\begin{itemize}
    \itemsep0em 
    \item $\gwsss$, a novel graph-based weakly-supervised method to jointly segment and classify \glspl{wsi} using readily available \gls{wsi}-level annotations. 
    \item A comprehensive evaluation of $\gwsss$ on 3 prostate cancer datasets for Gleason pattern segmentation and Gleason grading, and benchmarked against state-of-the-art \gls{wsi}-level weakly-supervised methods.
    \item Thorough generalizability quantification of $\gwsss$ on \textit{in-} and \textit{out-of-domain} cohorts in terms of segmentation and classification performance, uncertainty estimation, and calibration of neural network predictions.
\end{itemize}

A preliminary version of this work was presented as~\cite{anklin21}. Our substantial extensions herein include, 
(1) an improved $\gwsss$ method in terms of model architecture and automatic synthesis of node labels,  
(2) extensive evaluations on large cohorts of \glspl{wsi} (approximately 100$\times$), and (3) generalization assessment.

\section{Related work}
\label{sec:related_work}
\subsection{Weakly-supervised histopathology image classification}

Weakly-supervised classification of \glspl{wsi} has been mostly developed around \gls{mil}. In \gls{mil}, a \gls{wsi} is first decomposed into a ``bag'' of patches and are encoded by a \emph{neural encoder}, \eg a \gls{cnn}. Then, an \emph{aggregator} pools the patch embeddings to produce a slide-level representation for mapping to a class label via a \emph{neural predictor}. The \emph{aggregator} can be based on an attention mechanism weighing the importance of each patch, as in \cite{ilse2018,lu20b}, or as recently proposed, it can take the form of a transformer~\cite{myronenko21,shao21} or a \gls{gnn}~\cite{lee2022derivation}, enabling modeling inter-patch dependencies and global context. Differently, context can be modeled using multi-scale representations of \glspl{wsi}, either via multi-magnification patch embeddings~\cite{ho21, lipkova2022} or by learning to automatically select important regions, as proposed in~\cite{thandiackal22,kong2022efficient}. Despite the success of these approaches, they cannot directly be extended for semantic segmentation.

\subsection{Weakly-supervised histopathology image segmentation}

\gls{wsss} approaches in histopathology can be categorized by the type of supervision (or annotation), \eg point annotations, scribbles, or image-level labels, and the scale of operation, \eg patches, tiles, \glspl{tma}, or \glspl{wsi}. 
%
% Point supervision: Cell level
\cite{lee2020scribble2label,qu2020weakly,yen2020ninepins} utilized point annotations to segment cells and nuclei in histology patches.
%
% Scribble supervision: Patch level
\cite{bokhorst2018learning,ho21} used scribble annotations to segment tissue and tumor regions, respectively, at patch-level. Both the approaches used U-Net~\cite{ronneberger2015unet}, where \cite{ho21} leveraged concentric patches across multiple magnifications for including relevant context information, and \cite{bokhorst2018learning} modified the objective function to balance the contribution of the annotated pixels.
%
% Image supervision: Patch level
The majority of the \gls{wsss} methods in histopathology utilized image-level supervision and are limited to operate with patch annotations. 
\cite{xu14} proposed multiple clustered instance learning to process sliding patches for simultaneous grading and segmentation of colon \glspl{tma}.
\cite{jia17} trained a binary classifier for pixel-level predictions and afterward computed an image-level prediction from pixel labels via a softmax function. They optimized image prediction, such that pixel predictions were improved.
\cite{xucamel19} proposed CAMEL, a \gls{mil}-based label enrichment method. It split an image into latticed instances, generated instance labels, and assigned instance labels to corresponding pixels to enable supervised segmentation.
\cite{chan19} proposed HistoSegNet, which trained a \gls{cnn} to predict tissue types in a tile and used feature attribution to derive pixel-level predictions. It also employed a series of dedicated post-processing steps for prediction refinement.
\cite{zhang2021joint} used foreground proportion as the weak labels and combined a fully convolutional network and a graph convolutional network for tissue segmentation.
\cite{han2022multi} proposed a feature attribution-based model to generate pseudo labels, followed by a multi-layer pseudo-supervision network for segmenting tissue types.
As a main limitation, these methods cannot perform \gls{wsss} on \glspl{wsi} using only \gls{wsi} labels.
%
% Image supervision: TMA level
To perform \gls{wsss} beyond patch-level, \cite{silva2021weglenet} proposed WeGleNet, that scales to \glspl{tma}. It included a segmentation- and a global-aggregation layer to classify images during training, and upsampled pixel-level softmax activations during inference for image segmentation. However, the method cannot precisely delineate lesions and highlight multiple lesion occurrences. It also requires processing densely overlapping patches for fine segmentation, and cannot scale to \glspl{wsi}. 
In contrast, our $\gwsss$ can perform \gls{wsss} by leveraging image-level supervision, while efficiently scaling to \glspl{wsi} of arbitrary dimensions.

\subsection{Generalization quantification in histopathology}
Generalizability of \gls{cad} tools in histopathology is affected by domain-level biases, which are introduced due to numerous reasons, such as different staining protocols, manufacturing devices, materials, and scanning devices with respective color response~\cite{aubreville21}. Though generalizable tools, that are robust to domain shifts, are desired, it is challenging to model and detect the domain shifts in \gls{dl}. Nevertheless, several approaches have been proposed to reduce such domain shifts via data- and model-level adaptation. 

Data-level adaptation can be achieved via stain normalization~\cite{macenko09, vahadane16, stanisavljevic18, ren19}, color augmentation~\cite{tellez18, faryna21}, or stain invariant feature learning~\cite{otalora2019staining,yamashita2021learning}. 
Model-level adaptation is typically done via domain adversarial training~\cite{ganin16, aubreville20, hashimoto2020multi, tschuchnig2020generative}, which leverages target domain unlabeled data along with source domain data for modeling.

However, the aforementioned data- and model-level adaptation approaches do not exhaustively assess the generalization ability of their trained \gls{dl} models beyond task performance. In this case, accurate \emph{uncertainty estimation} and \emph{model calibration} are crucial to know \emph{when} to trust the model -- a task known to be challenging for neural networks that often provide over-confident predictions~\cite{guo17,lakshminarayanan17,fort19}. To the best of our knowledge, computational pathology research in these directions is scarce and remains unexplored.

\section{Methodology}
\label{sec:methodology}
In this section, we present $\gwsss$ for scalable \gls{wsss} of histopathology images. First, we transform a \gls{wsi} into a \gls{tg} representation, where nodes and edges of the graph denote tissue regions and their interactions, respectively (Section~\ref{sec:tg_construction}). Next, a \gls{gnn} contextualizes node embeddings characterizing tissue regions (Section~\ref{sec:node_embedding}), which are then processed by a \textit{graph classification head} for Gleason grading (Section~\ref{sec:wsi_classification}). Finally, we generate node-level pseudo labels using  feature attribution and a node selection strategy, which are used to train a \emph{node classification head}. The \emph{node-head} outputs the segmentation mask with pixel-level Gleason pattern assignment (Section~\ref{sec:segmentation}). An overview of the method is presented in Figure~\ref{fig:wholesight_block_diagram}.

\begin{figure*}[!t]
    \centering
    \includegraphics[width=0.9\linewidth]{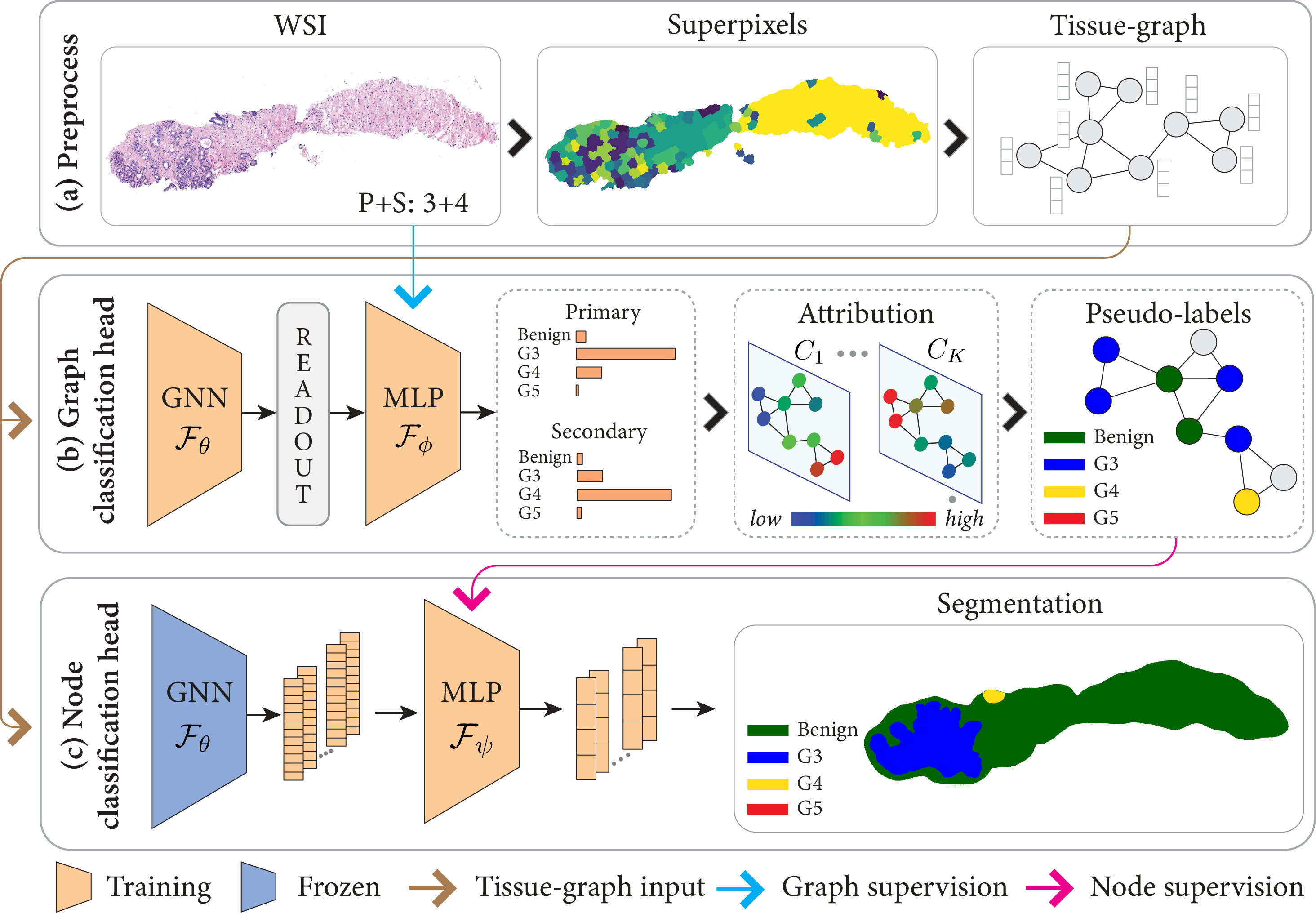}
    \caption{Overview of the proposed $\gwsss$ method. 
    (a) In the preprocessing step, a \gls{tg} is constructed to represent a \gls{wsi}, where the nodes and edges are defined by identifying superpixels and region adjacency connectivity, respectively. 
    (b) The \emph{graph classification head} classifies the \gls{tg} into primary and secondary Gleason patterns. Subsequently, a feature attribution technique and a node selection strategy derive node-level pseudo-labels.
    (c) The \emph{node classification head} learns on the pseudo-labels to classify the nodes, thereby resulting in the \gls{wsi} segmentation.
    }
    \label{fig:wholesight_block_diagram}
\end{figure*}

\subsection{Notation  and preliminaries}
We define a graph $G \in \mathcal{G}$ as $(V_G,E_G,H)$, where $V_G$ and $E_G$ denote the set of nodes and edges, respectively, $H \in \real^{|V| \times d}$ denotes $d$-dimensional node features (or denoted at node-level as $H_{v,.} := h(v) \in \real^d$), and $\mathcal{G}$ is the set of graphs. The neighborhood of a node $v \in V_G$ is denoted as $\mathcal{N}(v) := \{u \in V_G \; | \; (v,u) \in E_G \; \vee \; (u,v) \in E_G\}$.
We represent the cardinality of a set as $|.|$, \eg $|\mathcal{N}(v)|$ indicates the number of neighbors of $v$.

\glspl{gnn}~\cite{hamilton2020graph} are a class of neural networks that learn from graph-structured data. Specifically, \glspl{gnn} follow a two-step procedure to contextualize node features by including neighborhood node information. First, in an $\aggregate$ step, for each node $v \in V_G$, the neighboring node features $\mathcal{N}(v)$ are aggregated by a differentiable and permutation-invariant function. Next, in an $\update$ step, the current features of $v$ and the aggregated features of $\mathcal{N}(v)$ are processed by a differentiable operator to update the features of $v$. This procedure is repeated $T$ times, where $T$ is the number of \gls{gnn} layers.

In this work, we use the \gls{gin}~\cite{xu19}, 
where the $\aggregate$ step is a \textit{mean}-operator, and the $\update$ step includes a multi-layer perceptron ($\mlp$). Formally, a \gls{gin} layer is given as,
\begin{align}
h^{(t+1)}(v) &= \mlp \Big( h^{(t)}(v) + \frac{1}{|\mathcal{N}(v)|} \sum_{u \in \mathcal{N}(v)} h^{(t)}(u) \Big) 
\end{align}
$T$ \gls{gin} layers, denoted as $\mathcal{F}_\theta$, are stacked to acquire context information up to $T$-hops for each $v$. For graph classification, a fix-sized graph-level embedding $h_G$ is derived by pooling the node  embeddings $h^T(v), \; \forall v \in V_G$ by a $\readout$ step, \eg a \textit{mean}-operation. Subsequently, $h_G$ is mapped to target classes by a classifier network, $\mathcal{F}_\phi$. Similarly, for node classification, $h^T(v), \; \forall v \in V_G$ can be classified by a classifier network $\mathcal{F}_\psi$.

% learning and loss basics 
Formally, classification aims to predict target label $y \in \mathcal{K}$ for an input $x \in \mathcal{X}$, where $\mathcal{K}$ and $\mathcal{X}$ denote the set of classes and inputs, respectively. Given a set of sample pairs $\{ (x_i, y_i) \}_{i=1}^N$, where $N$ is the number of samples and $(x_i, y_i) \sim p(x,y)$, the data likelihood can be expressed as $p(Y|X, \theta, \phi) = \Pi_{i=1}^N p(y_i | x_i, \theta, \phi)$. The optimal parameters $(\hat{\theta}, \hat{\phi})$ are obtained by maximum likelihood estimation, or equivalently by minimizing the \gls{nll} $- \sum_{i=1}^N \log p(y_i | x_i, \theta, \phi)$. For graph classification, a sample pair is denoted as $(y_G, G), \; y_G \in \mathcal{K_G}, \; G \in \mathcal{G}$. For node classification, a sample pair is denoted as $(y_V, v), \; y_V \in \mathcal{K_V}, \; v \in \mathcal{V}$. For the task in this paper, the set of graph- and node-level classes are the same, \ie $\mathcal{K}:= \mathcal{K_G} = \mathcal{K_V}$. 

% calibration 
We further introduce the notion of model \emph{calibration}~\cite{nixon2019measuring}.
Intuitively, the probability of outcomes, \ie confidence scores, of a calibrated model should match its performance. For example, the samples predicted with an average confidence of 60\% by a model should have an average accuracy of 60\%. Formally, for a given network, $f: \mathcal{X} \rightarrow \mathcal{K}$, and $p(X, Y)$ a joint distribution over the data and the labels, $f(x)$ is said to be calibrated with respect to $p$ if, $\mathbb{E}_p[Y | f(X) = \beta] = \beta$, $ \forall \beta \in [0, 1]$. The \emph{calibration} can be visualized with a \emph{reliability diagram}~\cite{degroot83}. Namely, all the samples in the dataset are assigned to bins according to their predicted confidence scores. Then, the model accuracy is computed for the samples in each bin. The network performance is plotted against the binned confidence scores, where deviations from the diagonal represent uncalibrated bins.

\subsection{Preprocessing and tissue-graph construction} \label{sec:tg_construction}
First, we stain-normalize the input H\&E stained images using the method by~\cite{vahadane16}. It reduces appearance variability across images caused during tissue preparation, \ie different specimen preparation techniques, staining protocols, fixation characteristics, and imaging device characteristics~\cite{veta14, tellez19}. Then, we transform the normalized images into \glspl{tg} (Figure~\ref{fig:wholesight_block_diagram}(a)), where the nodes and the edges of a \gls{tg} denote tissue regions and inter-tissue interactions, respectively. Motivated by~\cite{bejnordi15,pati2022interpretation}, we consider superpixels as the visual primitives to encode tissue regions. Compared to rectangular patches, superpixels are more flexible to accommodate arbitrary shapes according to the local homogeneity of tissue. The homogeneity constraint also restricts the superpixels to span across multiple distinct structures and include different morphological regions.

\gls{tg} construction follows \cite{pati21}, where the prominent steps are, (1) detection of superpixels to define nodes $V_G$, (2) characterization of superpixels to define node features $H$, and (3) building graph topology to define edges $E_G$.
We adopt a two-step process to identify superpixels in a \gls{wsi}. First, we use \gls{slic}~\cite{achanta12} to produce over-segmented superpixels. Over-segmentation is conducted at a low magnification to capture homogeneous regions while offering a good compromise between granularity and smoothing-out noise. In the second step, the over-segmented superpixels are hierarchically merged according to their channel-wise color similarity at high magnification. Color similarity is quantified in terms of channel-wise 8-bin color histograms, mean, standard deviation, median, energy, and skewness. The resulting merged tissue regions form the nodes of the \gls{tg}. The merged superpixels denote morphologically meaningful homogeneous regions. Additionally,  merging reduces the node complexity of the \gls{tg}, thus enables the scaling of \gls{tg} to a large \gls{wsi} and contextualization to distant tissue regions.

We characterize the \gls{tg} nodes by morphology and spatial features. Considering the potentially arbitrary dimension of superpixels, we use a two-step process to derive morphology. First, we extract patches of size 144$\times$144 pixels from a superpixel, resize them to 224$\times$224 size, and encode them into 1280-dimensional features via MobileNetV2 network~\cite{sandler18} pre-trained on ImageNet~\cite{deng09}. Superpixel-level features are computed as the mean of the patch-level features. 
Next, we compute spatial features for each node by normalizing the superpixel centroids by the image dimensions. Normalization ensures the invariability of the spatial features to the varying dimensions of input \glspl{wsi}. 
Finally, we define the \gls{tg} edges by constructing a region adjacency graph topology~\cite{potjer96} using the spatial connectivity of superpixels. To this end, we assume that adjacent tissue regions biologically interact the most, and thus should be connected in a \gls{tg}.

\subsection{Contextualization of node embeddings} \label{sec:node_embedding}
Given a \gls{tg}, we learn discriminative node embeddings (see Figure~\ref{fig:wholesight_block_diagram}(b)) by using the node context information, \ie the tissue microenvironment and the inter-tissue interactions. Specifically, we use \gls{gin}~\cite{xu19} denoted as $\mathcal{F_{\theta}}$. Since \glspl{gnn} can operate on graphs of arbitrary and varying sizes, they allow to encode histopathology images represented in form of \gls{tg}s without needing tile-based processing. As the discriminative information of a node relies on its local sub-graph structures and can lie at different abstraction levels in the \gls{gnn}, we employ a Jumping Knowledge~\cite{xu18} strategy to utilize multi-level node representations. 
Namely, the final node-level embedding after $T$ \gls{gin}-layers is defined as,
\begin{align}
h^{(T)}(v) &= \concat(h^{(t)}(v), \; \forall t \in \{1, ..., T \}) 
\end{align}
where $\concat$ denotes a concatenation operation.

\subsection{\gls{wsi} classification} \label{sec:wsi_classification}
Following the contextualization of node features, a \textit{graph-classification head} classifies the \gls{tg} by using graph-level embeddings $h_G$ and graph/image-level supervision. To obtain a fix-sized $h_G$, we use a $\readout$ operation that averages the node embeddings $h^{(T)}(v), \; \forall v \in V_G$. Subsequently, $h_G$ is input to a multi-task classifier for primary and secondary Gleason grading. Specifically, the classifier includes two \glspl{mlp}, denoted as $F_{\phi} = \{F_{\phi_1}, F_{\phi_2}\}$, to individually predict the primary, \ie the worst Gleason pattern, and secondary, \ie the second-worst Gleason pattern, in the \gls{wsi}. Each \gls{mlp} solves a multi-class problem with $|\mathcal{K}|$ Gleason patterns, \ie benign, Grade 3, Grade 4, and Grade 5. The final Gleason grade is derived as the sum of the predicted primary and secondary patterns. $\mathcal{F_{\theta}}$ and $\mathcal{F_{\phi}}$ are optimized jointly by minimizing the weighted cross-entropy loss,
\begin{align}
\mathcal{L}_G = \lambda \mathcal{L}_{CE}(y_{G_{P}}, \hat{y}_{G_{P}}) + (1-\lambda)\mathcal{L}_{CE}(y_{G_{S}}, \hat{y}_{G_{S}}) 
\end{align}
where, $P$ and $S$ denote primary and secondary labels of ground truth $y_G$ and prediction $\hat{y}_G$, and $\lambda \in [0,1]$ is a hyper-parameter balancing the two terms.
Further, during training we introduce class-weights as $w := \{ \log(\frac{\sum_i N_i}{N_i}),\; i=\{1,...,|\mathcal{K}|\} \}$, where $N_i$ is the count of class-wise Gleason patterns in the training \glspl{wsi}. These weights take care of the class imbalance in Gleason grading by assigning a higher value to classes with lower frequency.

\subsection{Weakly supervised semantic segmentation} \label{sec:segmentation}

Nodes in a \gls{tg} identify superpixels, \ie morphologically homogeneous tissue regions. Since each Gleason pattern is characterized by \emph{distinct} morphological patterns, we assume that each tissue region, depicted by a node, includes a \emph{unique} Gleason pattern. Thus, the \gls{wsi} segmentation task is translated into classifying the nodes of the \gls{tg}. In presence of only image supervision, the node classification is achieved in two steps. First, pseudo-node labels are generated by using the image labels, and then, the pseudo labels are used to train a node classifier.

\textit{\underline{Pseudo node labels}:}
Following \gls{wsi} classification, a post-hoc \emph{feature attribution} technique is used to measure the importance of each node towards \gls{tg} classification. Specifically, we use $\graphgradcam$~\cite{pope19, jaume21}, an extension of $\gradcam$~\cite{selvaraju17} to operate with \glspl{gnn}. Given a graph $G$, $\graphgradcam$ produces class-wise node attribution maps, $A_k, \; \forall k \in \mathcal{K}$. These maps highlight the importance $\forall v \in V_G$ for classifying $G$ into $|\mathcal{K}|$, as shown in Figure~\ref{fig:wholesight_block_diagram}. Provided the importance scores for $v$ towards $|\mathcal{K}|$, it can be assumed that the label of $v$ is $k \in \mathcal{K}$, if the highest importance score corresponds to class $k$. At this stage, an \emph{argmax} operation across $A_k, \; \forall k \in \mathcal{K}$ can be considered to classify the nodes. However, such node labeling may be suboptimal, because,
\begin{itemize}
    \itemsep 0em
    \item Some nodes marginally contribute and bear low importance scores $\forall k \in \mathcal{K}$ for classifying a graph. However, an \emph{argmax} across the importance scores for a node greedily selects the class with the highest score, even though the node label is not ascertained.
    \item A node highly contributing towards the prediction of a class is not necessarily part of this class. For example, a node can bear high importance if it provides useful complementary information for tie-breaking or ruling out another class possibility. Formally, if the set of nodes $V_k \subset V$ has high importance scores for class $k$, the labels of $V_k$ are not ensured to be $k$. Even, the labels of $v \in V_k$ are not guaranteed to be the same.
    \item A class attribution map does not necessarily highlight all the nodes belonging to the class. Depending on the task complexity, a classifier may utilize only a subset of the informative nodes from a class to predict the graph label. Formally, if the set of nodes $V_k \subset V$ have high importance scores for class $k$, then $V_k$ may not include all the nodes in $\mathcal{V}_k \subset V$ that have the actual label $k$, \ie $V_k \subset \mathcal{V}_k$.
    \item In presence of several feature attribution techniques in literature, with different  underlying mechanisms, can produce different attribution maps~\cite{jaume21}. Thus, a single attribution technique may not be trusted for score-based node classification.
\end{itemize}

We, therefore, strategize to use the highlighted nodes by feature attribution as pseudo-labels to train a \textit{node-classifier}. For a graph $G$ with Gleason score $P$+$S$, $P, S \in \mathcal{K}$, we compute node importance scores $I_P$ and $I_S$, $\forall v \in V_G$ using $\graphgradcam$. As the scores by $\graphgradcam$ are unbounded, we normalize the scores using min-max. Then, we select the top $n\%$ nodes above a threshold $t$, denoted as $V_P$ and $V_S$, where $n$ and $t$ are hyperparameters tuned during training. It selects the most informative nodes for downstream node classification. For a node $v \in V_P$ and $v \in V_S$, we use $\arg\max(I_P(v), I_S(v))$ to ensure $V_P \cap V_S = \emptyset$. Finally, classes with the highest scores are assigned as pseudo labels $y_{\tilde{V}}$ to the nodes. Pursuing the process for all the \gls{tg}s in the dataset renders pseudo labels $Y_{\tilde{V}}$. 

\textit{\underline{Node classification}:}
$Y_{\tilde{V}}$ is used to train the \emph{node-classification head}, as shown in Figure~\ref{fig:wholesight_block_diagram}. Specifically for a graph $G$, we get the node embeddings $h^{(T)}(v), \, \forall v \in V_G$ using $\mathcal{F_{\hat{\theta}}}$, where $\hat{\theta}$ are the parameters of the \gls{gnn}. $\mathcal{F_{\hat{\theta}}}$ is frozen during node classification such that the same \gls{gnn} backbone is used for both segmentation and classification, thereby reducing the number of trainable parameters. $h^{(T)}(v), \, \forall v$ are processed by an \gls{mlp} $\mathcal{F_{\psi}}$ to predict $Y_{\tilde{V}}$. $\mathcal{F}_\psi$ is trained by optimizing a weighted multi-class cross-entropy objective. Similar to the graph classification, class-weights are defined as $w := \{ \log(\frac{\sum_i N_i}{N_i}),\; i=\{1,...,|\mathcal{K}|\} \}$, where $N_i$ is the number of annotated nodes of class $i$. The node-wise predicted class labels are used to obtain the final segmentation prediction. Noticeably, $\gwsss$ does not include any customized post-processing, unlike~\cite{chan19}, thus being applicable to various tissues and segmentation tasks.

Notably, the \textit{graph-} and the \textit{node-classification heads} address complementary  tasks for a graph $G$, \ie at graph-level and at node-level, respectively. Therefore, following the training of $\mathcal{F}_\psi$, we unfreeze $\mathcal{F}_\theta$, and jointly fine-tune $\hat{\theta}$ and $\hat{\psi}$ with a small learning rate. The complementarity of the tasks provides an additional informative signal to further improve the segmentation and classification performance of $\gwsss$.

\section{Experiments}
\label{sec:experiments}

\subsection{Datasets}
We evaluate our method on three datasets containing whole-slide prostate cancer needle biopsies for Gleason pattern segmentation and Gleason grading. Gleason patterns include grade 3 (G3)- moderately differentiated nuclei and poorly-formed cribriform glands, grade 4 (G4)- poorly differentiated nuclei and irregular masses, and grade 5 (G5)- less differentiated nuclei and lack or only occasional glands. Normal glands and non-epithelial tissues are labeled as benign (B). Gleason grade depicts the worst (\emph{primary}, P) and the second-worst (\emph{secondary}, S) Gleason patterns in a \gls{wsi}. Dataset details are as follows:

\begin{figure*}[t]
    \centering
    \includegraphics[width=0.87\linewidth]{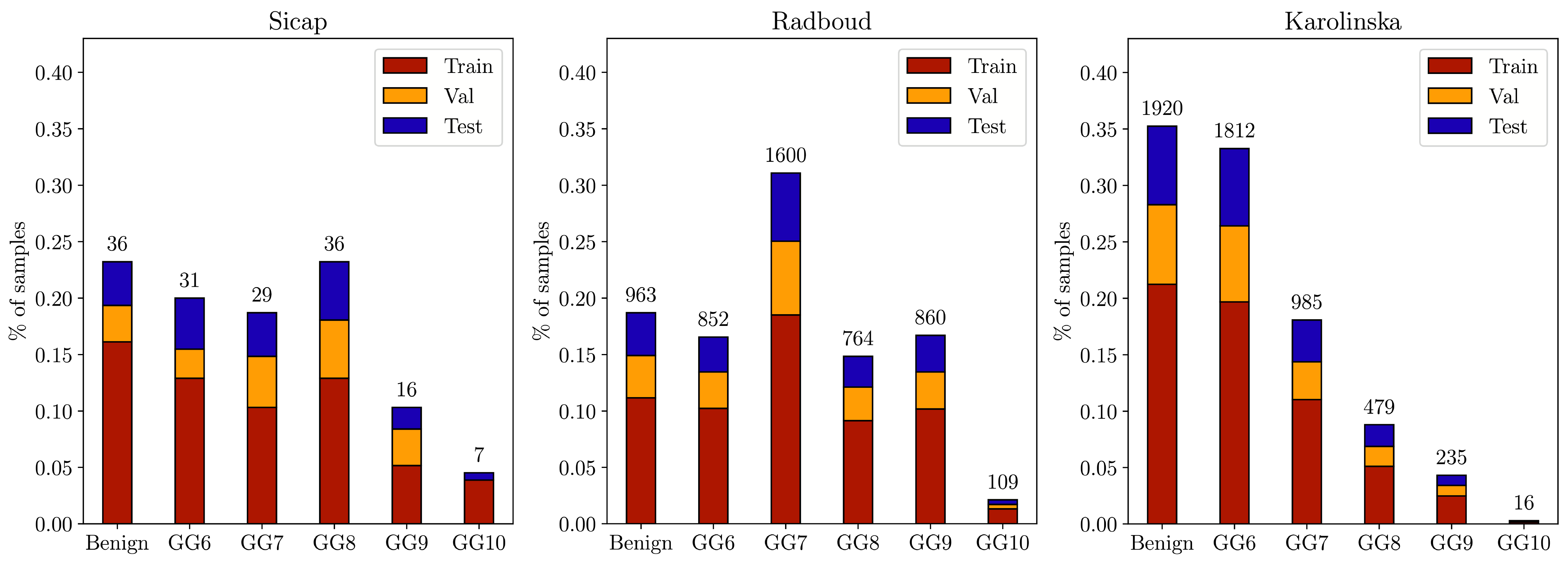}
    \caption{Gleason grade-wise data distribution across train, validation, and test in  Karolinska, Radboud and Sicap datasets.}
    \label{fig:data_distribution}
\end{figure*}

\textit{\underline{Sicap dataset}:} The dataset~\cite{silva20} contains 18,783 patches of size 512$\times$512 with \textit{complete} pixel-level annotations and slide-level Gleason grades for 155 \glspl{wsi} from 95 patients. The original slides and masks were reconstructed by stitching the patches. The \glspl{wsi} were scanned at 40$\times$ magnification by Ventana iS-can Coreo scanner and downsampled to 10$\times$ magnification. The slides were annotated by expert urogenital pathologists at the Hospital Clínico of Valencia, Spain. 

\textit{\underline{Radboud dataset}:}~\cite{bulten20} includes 5,759 needle biopsies from 1,243 patients at the Radboud University Medical Center, Netherlands.
The slides were scanned with a 3D Histech Panoramic Flash II 250 scanner at 20$\times$ magnification (resolution 0.24$\mu$m/pixel) and were downsampled to 10$\times$. Annotations include \gls{wsi} Gleason grades and noisy pixel-level Gleason pattern masks, released as part of the Prostate cANcer graDe Assessment (PANDA) challenge~\cite{bulten2022}. 
The masks were cleaned for segmentation using standard image manipulation techniques, \ie contextualized noise removal, hole filling, and edge smoothing. In absence of large public datasets with pixel-level annotated prostate cancer \glspl{wsi}, we used this dataset for developing and evaluating our method.

\textit{\underline{Karolinska dataset}:} The dataset~\cite{strom19} comprises of 5,662 core needle biopsies from 1,222 patients at various hospitals in Stockholm, Sweden.
The slides were scanned with a Hamamatsu C9600-12 and an Aperio Scan Scope AT2 scanner at 20$\times$ magnification with a pixel resolution of 0.45202$\mu$m and 0.5032$\mu$m, respectively. The biopsies were annotated by an expert uro-pathologist for Gleason grading. 

Each dataset is split into train, validation, and test in a ratio of 60\%, 20\%, and 20\% at Gleason grade level, using a random stratification that preserves the percentage of classes in each split. 
The dataset distributions and splits are displayed in Figure~\ref{fig:data_distribution}, which highlights the class-level imbalances.

\subsection{Implementation and evaluation}

We implemented $\gwsss$ using PyTorch~\cite{paszke19}, DGL~\cite{wang19}, and Histocartography~\cite{jaume21b}, and conducted experiments on NVIDIA Tesla P100 GPU and POWER9 CPU.

To develop the $\gwsss$ network, $\mathcal{F}_\theta$, $\mathcal{F}_\phi$, and $\mathcal{F}_\psi$ were designed by optimizing their respective hyperparameters. First, $\mathcal{F}_\theta$ and $\mathcal{F}_\phi$ were trained by using image-level labels, and then pseudo-node labels were created to train $\mathcal{F}_\psi$ to produce segmentation output. The number of \gls{gin} layers in $\mathcal{F}_\theta$ were optimized for the values $\{3, 4, 5\}$,  where the $\update$ function was defined as a 2-layer $\mlp$ with 64 hidden units and ReLU activations. $\mathcal{F}_\phi$ contains two heads for classifying \emph{primary} and \emph{secondary} Gleason grades, where each head consists of a 2-layer \gls{mlp} with 128 hidden units and ReLU activations. $\mathcal{F}_\psi$ contains a 2-layer \gls{mlp} with 128 hidden units and ReLU activations.

Considering the small size of the Sicap dataset, node-level augmentations were employed to augment the training dataset. Specifically, random node rotations $\{90, 180, 270\}$ degrees, and horizontal and vertical mirroring were used for augmenting the nodes. Batch size and learning rate were optimized from $\{4, 8, 16\}$ and $\{10^{-4}, 5\times10^{-4}, 10^{-3}\}$, respectively. Dropout layers with rates $0.2$, $0.5$ and $0.5$ were included in the \glspl{mlp} belonging to $\mathcal{F}_\theta$, $\mathcal{F}_\phi$, and $\mathcal{F}_\psi$, respectively. The pseudo-node labels were extracted for selection percentages in $\{5, 10, 15, 20\}$ and thresholds $\{0.5, 0.6, 0.7\}$. Following the hyperparameter tuning, ten $\gwsss$ models were trained with different network initializations. Validation weighted-F1 was used for model selection. The reported results correspond to the mean and standard deviation over these ten models.

\textit{\underline{Classification metrics}:}
Classification performance was measured by the weighted-F1 score of Gleason grade and the quadratic kappa score ($\kappa^2$) of ISUP grade~\cite{epstein05, epstein16}. ISUP is an alternate grading system whose correspondence with Gleason grading is defined as, Benign $\rightarrow$ ISUP-0, GG-(3+3) $\rightarrow$ ISUP-1, GG-(3+4) $\rightarrow$ ISUP-2, GG-(4+3) $\rightarrow$ ISUP-3, GG-8 $\rightarrow$ ISUP-4, and GG$\geq$9 $\rightarrow$ ISUP-5. $\kappa^2$ captures the degree of disagreement between the prediction and ground truth labels. For example, a grade 6 sample predicted as grade 10 is penalized more than predicting grade 7.

\textit{\underline{Segmentation metrics}:}
Segmentation performance was measured by Dice score. Given the imbalance of the Gleason patterns in the datasets, we also reported the per-pattern Dice score. 

\textit{\underline{Uncertainty metrics}:}
Following the work of \cite{gomariz21}, we evaluated the classification uncertainties in terms of Brier score $s_{\text{B}}$ (lower is better) and the \gls{nll} $s_{\text{NLL}}$ (lower is better) over a set of $N$ test samples, defined as,
\begin{align}
    s_{\text{B}} = \frac{1}{N} \sum_{n=1}^N \sum_{i=1}^{|\mathcal{K}|} (y_i - \hat{y_i})^2, \;\;\; s_{\text{NLL}} = - \frac{1}{N} \sum_{n=1}^N \sum_{i=1}^{|\mathcal{K}|} p(y_i) \log \hat{p}(y_i)
\end{align}

\textit{\underline{Calibration metrics}:}
Reliability diagrams provide an intuitive understanding of model calibration. To quantify these observations, we used the \gls{ece} metric~\cite{kumar18}, which computes the weighted average deviation of the confidence scores over all the bins, \ie
\begin{align}
    c_{\text{ECE}} = \sum_{b=1}^B \frac{N_b}{N} |\text{acc}(b) - \text{conf}(b)|
\end{align}
where $n_b$ is the number of samples in bin $b$, $\text{acc}(b)$ and $\text{conf}(b)$ are the accuracy and the average confidence of samples in $b$.

\subsection{Baselines}
We compared $\gwsss$ with state-of-the-art \gls{wsi} classification methods and two variants of $\gwsss$.

\textit{\underline{FSConv}:} We implemented the two-step method proposed by \cite{silva20} for \gls{wsi} classification. First, we extracted patches of size 256$\times$256 from \glspl{wsi} and classified them using FSConv+global-max pooling. The patches were labeled using the Gleason pattern masks, and patches with $>$90\% homogeneous pattern were selected for classifier training. During inference, dense patch predictions produced the output segmentation masks. An \gls{mlp} was trained on the Gleason grade percentages over the \gls{wsi} patches for Gleason grading.

% WholeSIGHT gradcam
\textit{\underline{$\gwsss$(Graph, $\graphgradcam$)}:} In comparison to $\gwsss$, this baseline contained only $\mathcal{F}_\theta$ and $\mathcal{F}_\phi$. It did not create or utilize pseudo labels, and the segmentation output was obtained by taking the \emph{argmax} over the class-wise $\graphgradcam$ attribution maps.

% WholeSIGHT combined 
\textit{\underline{$\gwsss$(Multiplex, NC)}:} This variant used both image- and pixel-level supervision during training and acts as the upper bound for $\gwsss$. As pixel-level annotations were available, $\mathcal{F}_\psi$ was trained using ground-truth node-level labels, instead of generating pseudo-node labels. The model consisted of the same  $\mathcal{F}_\theta$, $\mathcal{F}_\phi$, and $\mathcal{F}_\psi$ as the $\gwsss$ architecture. In this baseline, $\mathcal{F}_\theta$, $\mathcal{F}_\phi$, and $\mathcal{F}_\psi$ were trained jointly by optimizing a multi-task objective, \ie \gls{wsi}-level primary and secondary Gleason score prediction along with node-level Gleason pattern prediction. This variant of $\gwsss$ was proposed in our preliminary work, as described in~\cite{anklin21}.

\textit{\underline{Multiple Instance Learning (\gls{mil})}:} \gls{mil} methods are state-of-the-art for \gls{wsi} classification. In particular, we compared to ABMIL~\cite{ilse2018}, which used an attention mechanism to aggregate patch embeddings into a fix-sized \gls{wsi} embedding that was fed to a classifier for Gleason grading.  We also included CLAM~\cite{lu20b}, a method built on ABMIL by including an additional constrain to cluster similar patch embeddings.
% -constrained attention \gls{mil} approach designed for \gls{wsi} classification.
Our experiments followed the public implementations~\footnote{CLAM publicly available code: https://github.com/mahmoodlab/CLAM} with adjustments to enable multi-task classification.

For all the baselines, hyper-parameters are thoroughly tuned to use the best learning rate and batch size, if applicable. Subsequently, ten models were re-trained from scratch with the optimal parameters. We report the mean and standard deviation over these runs for each experiment.

\subsection{\gls{wsss} performance analysis}
We studies the classification and segmentation performance of $\gwsss$ and the competing methods by independently training and testing them on Sicap, Radboud, and Karolinska.

% Sicap results 
\begin{table*}[t]
\centering
\caption{Classification and segmentation results on \textbf{Sicap} dataset. The best performances for using image-level supervision are highlighted in \textbf{bold}.}
\begin{tabular}{l|l@{~~}l@{~~}l@{~~}l@{~~}l@{~~}l@{~~}l@{~~}l}
\toprule
\parbox[t]{3mm}{\multirow{2}{*}[0ex]{\rotatebox[origin=c]{90}{Annot.}}} & & \multicolumn{4}{c}{per-class Dice}  & avg. Dice & GG wF1 & ISUP $\kappa^2$        \\
\cmidrule(lr){3-6}
 & Method & Benign & Grade3 & Grade4 & Grade5 & & & \\
\midrule
\parbox[t]{3mm}{\multirow{2}{*}[0ex]{\rotatebox[origin=c]{90}{$\mathcal{C}$}}} & FSConv~\cite{silva20} & 65.7\std{0.5} & 24.4\std{1.4} & 29.0\std{1.4} &  8.4\std{0.6} & 31.9\std{0.5} & 48.7\std{3.4} & 50.9\std{3.4} \\
& $\gwsss$ (Multiplex, NC) & 92.5\std{0.3} & 35.4\std{2.3} & 51.6\std{2.0} & 11.1\std{2.2} & 47.6\std{2.1} & 59.6\std{4.1} & 84.6\std{3.2} \\
% & (Multiplex, NC) &  &  &  &  &  &  &  \\
\midrule

\parbox[t]{3mm}{\multirow{6}{*}[0ex]{\rotatebox[origin=c]{90}{\shortstack[c]{$\mathcal{W}$}}}}
& ABMIL\cite{ilse2018} & - & - & - & - & - & 50.2\std{6.3} & 67.8\std{5.2} \\

& CLAM\cite{lu20b} & - & - & - & - & - & 51.4\std{5.5} & 75.2\std{4.7} \\

& $\gwsss$ & 64.4\std{6.1} & 23.1\std{2.0} & 32.8\std{6.9} & 3.7\std{1.0} & 31.0\std{2.7} & 53.3\std{5.3} & 81.9\std{6.7} \\
& (Graph, $\graphgradcam$) &  &  &  &  &  &  &  \\

& $\gwsss$ & \textbf{67.6\std{0.5}} & \textbf{28.6\std{0.3}} & \textbf{49.1\std{0.4}} & \textbf{5.0\std{0.4}} & \textbf{37.6\std{0.3}} & \textbf{58.6\std{6.2}} & \textbf{89.0\std{1.2}} \\
& (Graph + Pseudo nodes, Node class.) &  &  &  &  &  &  &  \\
\bottomrule
\end{tabular}
\label{tab:sicap_results}
\end{table*}

% Radboud results 
\begin{table*}[t]
\centering
\caption{Classification and segmentation results on \textbf{Radboud} dataset. The best performances for using image-level supervision are highlighted in \textbf{bold}.}
\begin{tabular}{l|l@{~~}l@{~~}l@{~~}l@{~~}l@{~~}l@{~~}l@{~~}l}
\toprule
\parbox[t]{3mm}{\multirow{2}{*}[0ex]{\rotatebox[origin=c]{90}{Annot.}}} & & \multicolumn{4}{c}{per-class Dice}  & avg. Dice & GG wF1 & ISUP $\kappa^2$        \\
\cmidrule(lr){3-6}
 & Method & Benign & Grade3 & Grade4 & Grade5 & & & \\
\midrule
\parbox[t]{3mm}{\multirow{2}{*}[0ex]{\rotatebox[origin=c]{90}{$\mathcal{C}$}}} & FSConv~\cite{silva20} & 84.3\std{0.1} & 53.3\std{0.5} & 62.5\std{0.3} &  36.9\std{0.5} & 59.2\std{0.1} & 45.9\std{1.3} & 69.4\std{0.6} \\

& $\gwsss$ (Multiplex, NC) & 91.5\std{0.1} & 63.9\std{0.4} & 66.2\std{0.3} & 36.8\std{1.2} & 64.6\std{0.4} & 68.9\std{0.9} & 83.8\std{0.9} \\
\midrule

\parbox[t]{3mm}{\multirow{6}{*}[0ex]{\rotatebox[origin=c]{90}{\shortstack[c]{$\mathcal{W}$}}}}

& ABMIL\cite{ilse2018} & - & - & - & - & - & 59.3\std{1.7} & 79.7\std{1.2} \\

& CLAM\cite{lu20b} & - & - & - & - & - & 60.3\std{1.6} & 80.2\std{1.5} \\

& $\gwsss$ & 71.3\std{2.2} & 26.9\std{1.0} & 24.9\std{1.2} & 15.1\std{0.5} & 34.6\std{0.6} & 66.0\std{1.0} & 82.2\std{0.5} \\
& (Graph, $\graphgradcam$) &  &  &  &  &  &  &  \\

& $\gwsss$ & \textbf{75.9\std{0.3}} & \textbf{32.9\std{1.0}} & \textbf{29.1\std{1.5}} & \textbf{19.6\std{0.7}} & \textbf{39.4\std{0.3}} & \textbf{67.9\std{0.3}} & \textbf{83.0\std{0.2}} \\
& (Graph + Pseudo nodes, Node class.) &  &  &  &  &  &  &  \\
\bottomrule
\end{tabular}
\label{tab:radboud_results}
\end{table*}

% Karolinska results 
\begin{table*}[t]
\centering
\caption{Classification results on \textbf{Karolinska} dataset. The best performances for using image-level supervision are highlighted in \textbf{bold}.}
\begin{tabular}{l|l@{~~}l@{~~}l}
\toprule
\parbox[t]{3mm} & & GG wF1 & ISUP $\kappa^2$        \\
\midrule
\parbox[t]{3mm}{\multirow{3}{*}[0ex]{\rotatebox[origin=c]{90}{\shortstack[c]{$\mathcal{W}$}}}}
& ABMIL~\cite{ilse2018} & 65.0\std{2.0} & 79.1\std{1.2} \\

& CLAM\cite{lu20b} & 63.6\std{2.5} & 77.6\std{2.0} \\

& $\gwsss$ (Graph) & \textbf{70.5\std{0.6}} & \textbf{80.2\std{0.7}} \\
\bottomrule
\end{tabular}
\label{tab:karolinska_results}
\end{table*}

\textit{\underline{Performance analysis}:}
% introduce table 
Table~\ref{tab:sicap_results} presents the results on Sicap. The analyses are grouped into two supervision settings, \ie \emph{complete} ($\mathcal{C}$) and \emph{weak} ($\mathcal{W}$). Setting-$\mathcal{C}$ utilizes both image- and pixel-level annotations, whereas, Setting-$\mathcal{W}$ only uses image-level labels. $\gwsss$ reached $37.6\%$ average Dice score, which significantly outperforms $\gwsss$ (Graph, $\graphgradcam$) by $+6.6\%$ in absolute. $\gwsss$ (Multiplex, NC), that acts as the upper bound, produced a significant gain in segmentation compared to $\gwsss$. The per-class Dice scores indicate that the benign patterns that constitute most tissue areas have a high detection rate compared to less occurring Gleason patterns. For the classification task, $\gwsss$ outperformed ABMIL and CLAM, both in terms of Gleason grade weighted-F1 and ISUP $\kappa^2$. Notably, 

Table~\ref{tab:radboud_results} presents the results on Radboud. $\gwsss$ rendered an absolute gain of $+4.8\%$ in average Dice score over $\gwsss$ (Graph, $\graphgradcam$). This confirms the utility of pseudo-node labels for superior segmentation. Similar to the observations on Sicap, benign patterns had a high detection rate, followed by G$3$, G$4$, and G$5$ patterns. As Radboud dataset includes more G$5$ patterns than Sicap, we observed a significant gain in detecting high-grade Gleason patterns. For the classification task, the observations were also consistent with the observations on Sicap. 

Table~\ref{tab:karolinska_results} presents the results on Karolinska. In absence of ground truth pixel-level annotations, the segmentation performances could not be computed. $\gwsss$ (Graph) outperformed the baselines in terms of classification performance. 

The observations across Table~\ref{tab:sicap_results}, \ref{tab:karolinska_results} and \ref{tab:karolinska_results} conclude that, jointly optimizing classification and segmentation objectives provide complementary information to improve the overall classification performance, \ie $\gwsss$ (Multiplex) $>$ $\gwsss$ $>$ $\gwsss$ (Graph).

\begin{table*}[!h]
\centering
\caption{Classification and segmentation results on Radboud, Karolinska, and Sicap datasets for models trained using both Radboud and Karolinska datasets.}
\begin{tabular}{l|l@{~~}l@{~~}l@{~~}l@{~~}l@{~~}l@{~~}l@{~~}l@{~~}l}
\toprule
\parbox[t]{3mm}{\multirow{2}{*}[0ex]{\rotatebox[origin=c]{90}{Annot.}}} & & \multicolumn{3}{c}{Radboud}  & \multicolumn{2}{c}{Karolinska} & \multicolumn{3}{c}{Sicap} \\
\cmidrule(lr){3-5}
\cmidrule(lr){6-7}
\cmidrule(lr){8-10}
 & Method & avg. Dice & GG wF1 & ISUP $\kappa^2$ & GG wF1 & ISUP $\kappa^2$ & avg. Dice & GG wF1 & ISUP $\kappa^2$\\
\midrule
\parbox[t]{3mm}{\multirow{2}{*}[0ex]{\rotatebox[origin=c]{90}{$\mathcal{C}$}}} & FSConv~\cite{silva20} & 59.2\std{0.1} & 45.9\std{1.3} & 69.4\std{0.6} &  34.5\std{1.1} & 40.1\std{1.3} & 49.5\std{0.4} & 52.1\std{2.1} & 53.8\std{1.7}\\

& $\gwsss$ (Multiplex, NC) & 64.5\std{0.3} & 69.0\std{1.0} & 83.6\std{0.9} & 71.2\std{0.7} & 82.5\std{1.3} & 60.0\std{0.5} & 65.5\std{2.5} & 85.6\std{2.8} \\
\midrule

\parbox[t]{3mm}{\multirow{6}{*}[0ex]{\rotatebox[origin=c]{90}{\shortstack[c]{$\mathcal{W}$}}}}

& ABMIL\cite{ilse2018} & - & 57.6\std{2.3} & 73.8\std{2.3} & 65.5\std{1.3} & 77.3\std{2.8} & - & 56.4\std{2.7} & 75.0\std{7.5}\\

& CLAM\cite{lu20b} & - & 61.7\std{2.1} & 78.6\std{1.3} & 69.3\std{1.3} & \textbf{82.8\std{1.0}} & - & 53.1\std{3.8} & 74.6\std{4.2}\\

& $\gwsss$ & 34.6\std{0.6} & 66.0\std{1.0} & 82.2\std{0.5} & 69.2\std{0.9} & 80.3\std{0.9} & 30.4\std{1.0} & 65.1\std{2.3} & 86.1\std{2.5}\\
& (Graph, $\gradcam$) &  &  &  &  &  &  &  & \\

& $\gwsss$ & \textbf{44.6\std{0.2}} & \textbf{66.2\std{0.1}} & \textbf{82.9\std{0.1}} & \textbf{70.2\std{0.1}} & 81.3\std{0.1} & \textbf{42.0\std{0.3}} & \textbf{65.2\std{0.1}} & \textbf{86.6\std{0.1}} \\
& (Graph + Pseudo nodes, Node class.) &  &  &  &  &  &  &  &  \\

\bottomrule
\end{tabular}
\label{tab:panda_results}
\end{table*}

\begin{figure*}[!ht]
    \centering
    \includegraphics[width=0.99\linewidth]{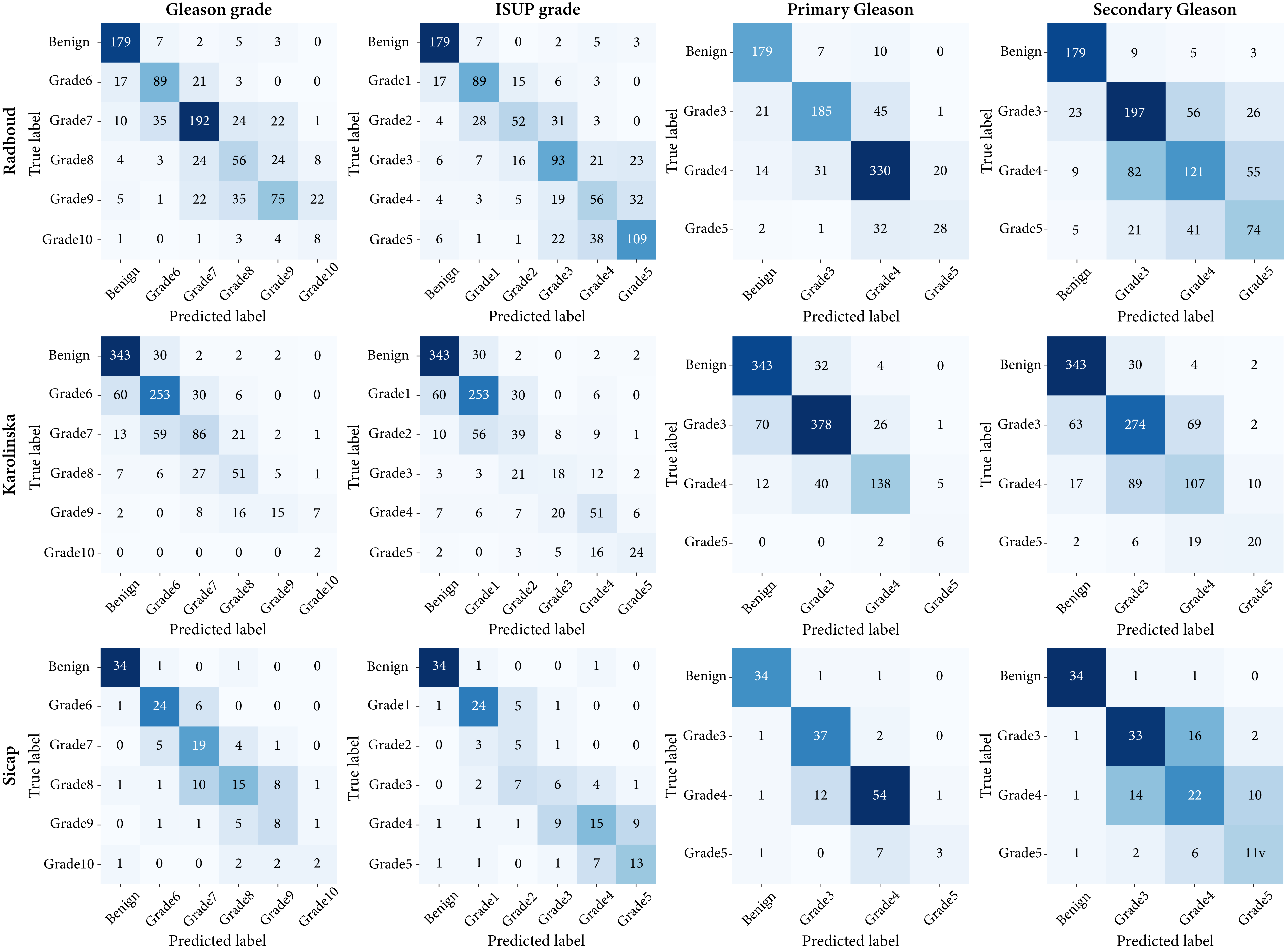}
    \caption{Confusion matrices for Gleason grading, ISUP grading, primary- and secondary Gleason classification on Radboud, Karolinska, and Sicap datasets for with the best $\gwsss$ model trained using Radboud and Karolinska training datasets.}
    \label{fig:classification_confusion_matrix}
\end{figure*}

\subsection{Generalization: performance, uncertainty, and calibration}
We studied the generalization ability of $\gwsss$ following a modified training setting. Specifically, we used Radboud and Karolinska training \glspl{wsi} for model training. Thus, the training set encompassed better sample variability and diagnostically more challenging cases than the standalone training counterparts on individual datasets. Testing was performed individually on  Radboud and Karolinska test \glspl{wsi}, herein studying the \emph{in-domain} performance. Further, we tested on the entire Sicap dataset, which consisted of \emph{out-of-domain} \glspl{wsi}.

\begin{figure*}[!h]
    \centering
    \includegraphics[width=0.97\linewidth]{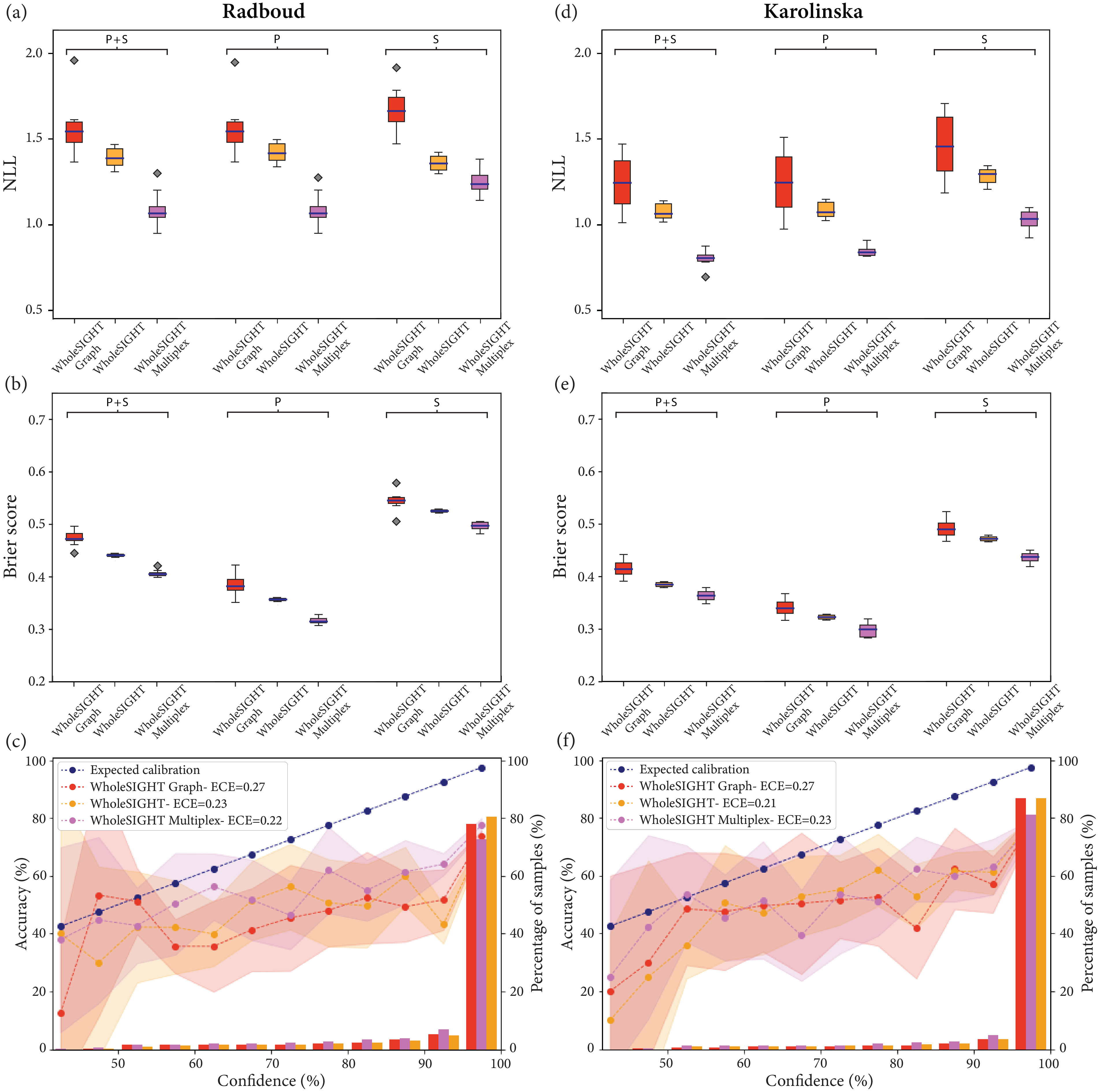}
    \caption{Uncertainty and model calibration analysis of $\gwsss$, $\gwsss$ (Graph, $\gradcam$), and $\gwsss$ (Multiplex, NC) models for Radboud (a, b, c) and Karolinska (d, e, f) datasets. (a, d) and (b, e) present NLL (lower is better) and Brier scores (lower is better), respectively.
    (c, f) present reliability diagrams of the primary Gleason classification head. Expected calibration (blue) highlights a perfectly calibrated model.  Calibrations of $\gwsss$, $\gwsss$ (Graph, $\gradcam$), and $\gwsss$ (Multiplex, NC) are in orange, red, and purple, respectively, along with the number of samples (in $\%$) in each bin.}
    \label{fig:uncertainty}
\end{figure*}

\textit{\underline{Performance analysis}:}
Table~\ref{tab:panda_results} compared the classification performance of $\gwsss$ and the competing baselines. In terms of the weighted-F1 score on the \emph{in-domain} test set, $\gwsss$ outperformed ABMIL and performed better or comparable to CLAM. Similar patterns were also observed for ISUP $\kappa^2$. However, the variances of the classification for $\gwsss$ is consistently lower than ABMIL and CLAM.
When tested on the \emph{out-of-domain} Sicap dataset, $\gwsss$ achieved significantly better classification than ABMIL and CLAM.

Confusion matrices for the best Gleason grading, ISUP grading, primary- and secondary classification with $\gwsss$ are presented in Figure~\ref{fig:classification_confusion_matrix} on the three datasets. It can be observed that most misclassifications lie close to the diagonal. Majority of the confusion occurred between GG$6$ and GG$7$, \ie GG$(3+3)$ versus GG$(3+4)$ and GG$(4+3)$. Such ambiguity is prevalent among pathologists, as shown in~\cite{ozkan2016interobserver,salmo2015audit}.
High-grade Gleason grading better on Radboud than Karolinska due to more number of high-grade samples in Radboud. Primary- and secondary classification weighted-F1 for Radboud, Karolinska and Sicap were 79.3\%, 81.7\%, 81.6\% and 62.5\%, 69.7\%, 64.5\%, respectively. This indicated that identifying secondary Gleason pattern is more challenging. 
%
% segmentation
Table~\ref{tab:panda_results} also presents the generalizability assessment of segmentation. $\gwsss$ consistently performed better than $\gwsss$ (Graph, $\graphgradcam$). Noticeably, the Dice scores on Radboud and Sicap datasets improved over the segmentation results in Table~\ref{tab:radboud_results}  and \ref{tab:sicap_results} by 5.2\% and 4.4\% for $\gwsss$. It can be reasoned to the usage of more training \glspl{wsi}, which indicate that \gls{wsss} can be improved by utilizing more weak supervision.  

\textit{\underline{Uncertainty analysis}:}
Figure~\ref{fig:uncertainty} presents the classification uncertainty analysis of $\gwsss$, $\gwsss$ (Graph, $\gradcam$), and $\gwsss$ (Multiplex, NC), in terms of \gls{nll} and Brier score, on Radboud and Karolinska datasets. $\gwsss$ (Multiplex, NC) rendered a significantly lower \gls{nll} than $\gwsss$ across all datasets for primary, secondary, and Gleason grade (P+S) classification. 
Noticeably, the \gls{nll} and Brier scores were consistently higher for predicting the secondary Gleason patterns than the primary patterns. This resonates with the fact that identifying secondary patterns is more challenging with higher ambiguity.

\textit{\underline{Model calibration analysis}:}
A model with good uncertainty estimate should be well-calibrated, \ie the model confidence should be close to the model performance. Figure~\ref{fig:uncertainty} presents the reliability diagrams of the primary classification head on Karolinska and Radboud datasets. $\gwsss$ showed consistently better calibration than  $\gwsss$ (Graph, $\gradcam$) and similar calibration with respect to $\gwsss$ (Multiplex, NC). \gls{ece} also metric quantitatively supported this observation. However, we observed that still all models remains over-confident as the model accuracies over the confidence bins remained lower than the expected calibration (in blue).

\subsection{Qualitative analysis}
We qualitatively analyze the results of $\gwsss$ by (1) visualizing overlaid segmentation masks on \glspl{wsi}, (2) analyzing the \gls{tsne}~\cite{van2008visualizing} node embeddings, and (3) correlating the segmentation outputs with pathological reasonings.

\begin{figure*}[!t]
    \centering
    \includegraphics[width=0.8\linewidth]{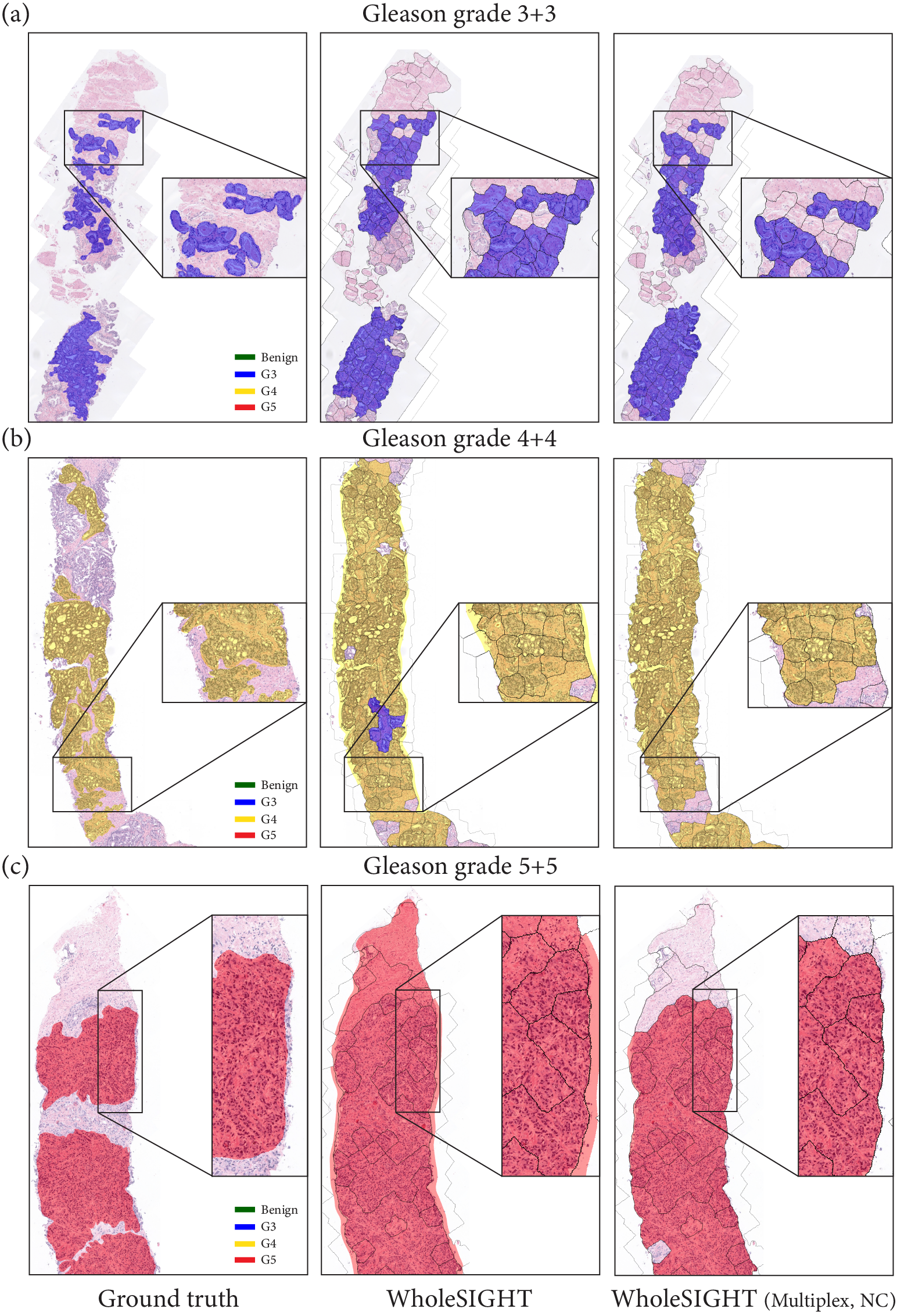}
    \caption{
        Sample segmentation maps from Sicap dataset. Ground truth is shown on the left, $\gwsss$ predictions in the middle, and $\gwsss$(Multiplex, NC) on the right. Tissue regions, \ie \gls{tg} nodes, are represented by black overlay. (a, b, c) display GG(3+3), GG(4+4), and GG(5+5) samples, respectively.
        For better visualization, benign areas are not highlighted in the segmentation maps.}
    \label{fig:qualitative_results}
\end{figure*}

\begin{figure*}[!t]
    \centering
    \includegraphics[width=0.95\linewidth]{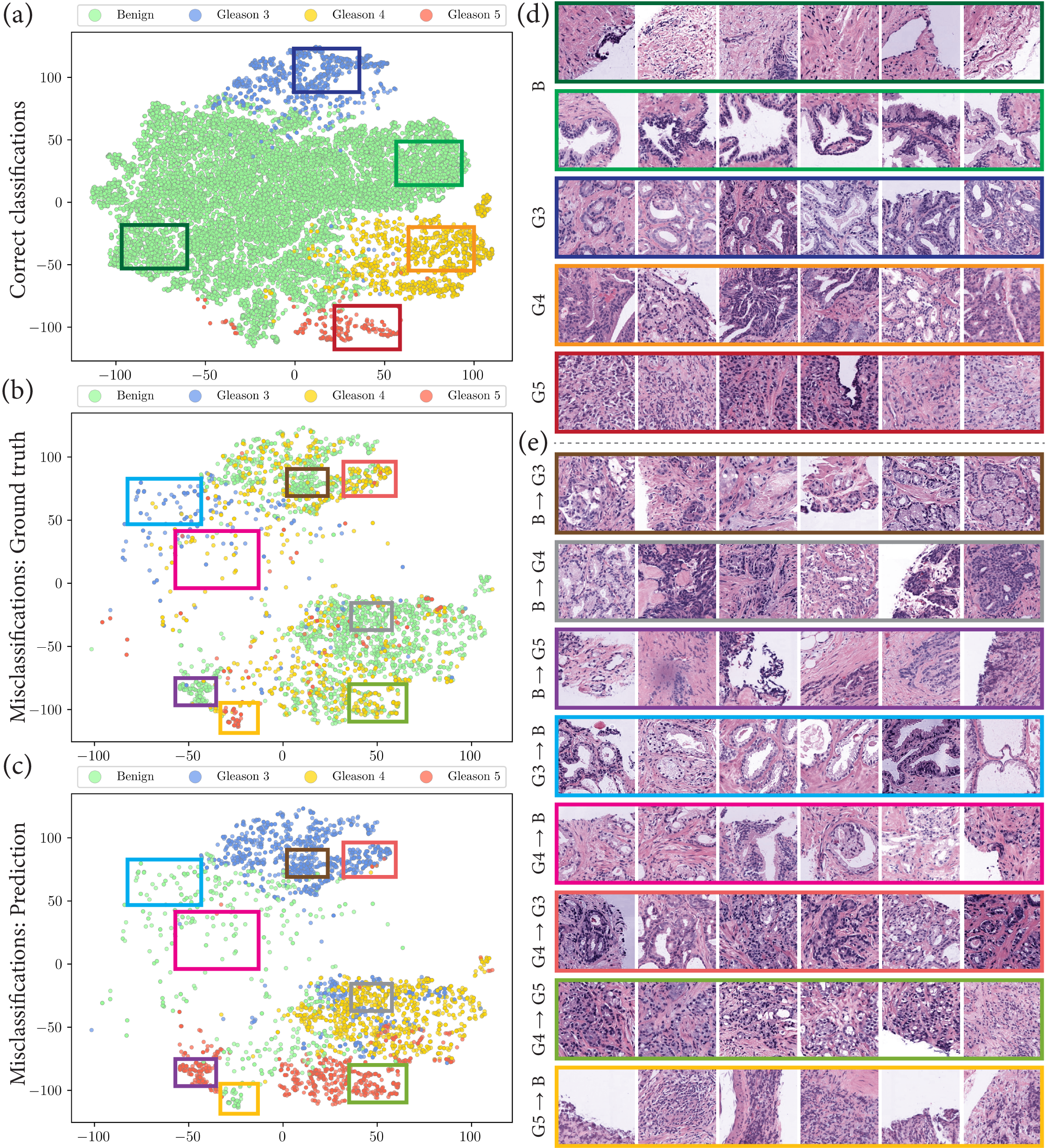}
    \caption{t-SNE visualization of tissue-graph node embeddings and example patches from several regions on the two-dimensional t-SNE feature space for Sicap dataset. (a) t-SNE visualization of the correctly classified nodes. (b) and (c) display the t-SNE visualization of misclassified nodes, where (b) and (c) highlight the ground truth and predicted node labels, respectively. (d) and (e) demonstrate square patches of size $224\times224$ at $10\times$ magnification cropped around the node centroids selected from different regions on the t-SNE embedding space. (d) and (e) highlight the correctly and incorrectly classified node patches, respectively. The labels of the patches in (e) are formatted as $Y \rightarrow \hat{Y}$, where $Y$ and $\hat{Y}$ denote the ground truth and the predicted class label. The colored rectangles around the patches in (d) and (e) correspond to respective colored rectangles in (a), (b), and (c).}
    \label{fig:tsne}
\end{figure*}

\textit{\underline{Segmentation mask visualization}:}
Figure~\ref{fig:qualitative_results} demonstrates segmentation predictions obtained with $\gwsss$ and $\gwsss$(Multiplex, NC) on Sicap dataset. We can observe that $\gwsss$ correctly delineates the cancerous regions in the \glspl{wsi}. Zooming into different regions conclude that the tissue regions of \gls{tg}, \ie the nodes of \gls{tg}, (outlined in black in Figure~\ref{fig:qualitative_results}) encode meaningful units of \emph{homogeneous} tissue.
It substantiates the relevance of using \gls{tg} representations for segmenting tissue regions into Gleason patterns. We further notice that $\gwsss$, in a few cases, predicts benign regions adjacent to cancerous patterns as cancerous. For example, the benign region, primarily consisting of stroma, in Figure~\ref{fig:qualitative_results}(c) is predicted as G5. We argue that these false positive detections do not inhibit the applicability of the method, as neighboring cancerous regions are correctly detected. In a few other cases, $\gwsss$ correctly detects missed cancerous regions in the ground truth annotations. For instance, in Figure~\ref{fig:qualitative_results}(b), the missing G4 region in the upper part of the \gls{wsi} is correctly identified.

% FP addressed with Combined 
Comparing $\gwsss$ with $\gwsss$ (Multiplex, NC), we observe that several false positives are removed, \eg in Figure~\ref{fig:qualitative_results}(a), thus offering more accurate segmentation. However, the improvements by $\gwsss$ (Multiplex, NC) are achieved at the cost of training with pixel-level annotations that are hardly available in real-world practice. Thus, $\gwsss$ appears to be an appealing compromise between segmentation performance and annotation requirement.

\textit{\underline{Visualizing \gls{tsne} feature space}:}
A \gls{tsne} visualization of the learned tissue-level embeddings is demonstrated in Figure~\ref{fig:tsne} for Sicap. \gls{tsne} projects the \gls{gnn} node embeddings onto a two-dimensional feature space, allowing to analyze the connection between node embeddings and the Gleason pattern distribution.

Figure~\ref{fig:tsne}(a) displays the \gls{tsne} feature space for the \emph{correctly} classified nodes, which highlights demarcated clusters for each Gleason pattern. The large cluster of benign nodes indicates the variability of the benign tissue. Several patches from each Gleason pattern cluster are presented in Figure~\ref{fig:tsne}(d). We can observe the reduced nuclei differentiation across the patches from benign to Gleason grade 5. Further, Figure~\ref{fig:tsne}(b) and (c) display the \gls{tsne} feature space for the misclassified nodes. Specifically, Figure~\ref{fig:tsne}(b) presents the ground truth node labels, and Figure~\ref{fig:tsne}(c) the predicted node labels. Different embedding locations are further selected and highlighted by different colored rectangles and put in relation with corresponding patches to indicate the inter-class ambiguities, as demonstrated in Figure~\ref{fig:tsne}(e). For example, the first row in Figure~\ref{fig:tsne}(e) showcases patches that are benign but are predicted as G3. We can visually compare these patches with the G3 patches in the third row of Figure~\ref{fig:tsne}(d). Similar ambiguities between other pairs of Gleason patterns are also included in Figure~\ref{fig:tsne}(e).

\textit{\underline{Interpreting model outcomes via predicted segmentations}:}
Predicted segmentations provide human-understandable \emph{interpretability} maps. For researchers, the segmentations allow to, (1) identify morphological patterns responsible for \gls{wsi} classification, (2) analyze failure cases by inspecting pixel-level predictions, and ultimately (3) better understand the model behavior towards biomarker discovery. For pathologists, they assist to, (1) put in relation the predicted \gls{wsi}-level Gleason scores and the highlighted pixel-level Gleason patterns, (2) confirm that the morphology of the identified cancerous regions aligns with pre-established diagnostic criteria.

Additionally, in the perspective of developing \gls{ai}-assisted human-in-the-loop tools, a Gleason grading system that can simultaneously \emph{classify} and \emph{segment} \glspl{wsi} is closer to the latest pathological standards.  Indeed, recent revisions of the Gleason grading system~\cite{epstein16} emphasized the importance of reporting the percentage of each grade for better patient stratification and treatment selection~\cite{cheng07, huang14, choy16, sharma20}. These percentages can be trivially derived from the predicted segmentation maps by counting the number of pixels belonging to each pattern. Naturally, such information is not available in mere \gls{wsi} classification systems. Reporting per-grade percentage is particularly important in ambiguous and borderline cases. For instance, consider two patients with Gleason score 3+4. When a small percentage of pattern-4 is present, \eg 10\%, the case can be considered as an intermediate risk cancer where active patient surveillance is enough~\cite{amin14}. However, a larger secondary pattern may require specific treatments. Reporting percentages of each grade allows us to discriminate between these two scenarios easily.

% example 2: In between two grades 
Similarly, consider a Gleason score 4+3 with a small secondary Gleason pattern, \eg 90\% and 10\% area for primary and secondary patterns, respectively. This case will be scored as 4+3, even though it is close to a score of 4+4, which would lead to a different treatment protocol. By explicitly reporting the Gleason pattern percentages, such corner cases can be avoided.

\section{Conclusion}
\label{conclusion}

Accurate delineation of patterns in whole-slide histopathology images typically demands pixel-level annotations, which are hard to acquire in a real-world scenario. Nonetheless, the semantic segmentation of diagnostically relevant patterns is crucial for disease diagnosis and treatment selection. To this end, we proposed a novel weakly-supervised semantic segmentation method, $\gwsss$, that can segment the relevant patterns of interest in histopathology images by leveraging only image-level supervision. To our knowledge, $\gwsss$ is the first weakly-supervised semantic segmentation method that can operate in an end-to-end manner on histopathology images of arbitrary shape and size.
We evaluated our proposed method on three publicly available prostate needle biopsy datasets for Gleason grade classification and Gleason pattern segmentation. On comparing state-of-the-art methods for histopathology applications, we demonstrated the classification and segmentation superiority of $\gwsss$.
Additionally, $\gwsss$ is a modular approach that can utilize both image-level and pixel-level supervision to simultaneously perform image classification and segmentation tasks.
Though we have evaluated our method for H\&E stained prostate cancer needle biopsies, the technology is easily extendable to other tissue types, imaging techniques, and image modalities.

\bibliographystyle{IEEEtran}
\bibliography{main}

\end{document}